\documentclass[aps,prl,twocolumn,groupedaddress,superscriptaddress,amsfonts,amssymb,amsmath,floatfix,citeautoscript]{revtex4-1} %
\usepackage{graphicx} 
\usepackage[centering,hmargin=24mm,tmargin=24mm,bmargin=24mm]{geometry}
\usepackage{amsmath}
\usepackage{hyperref} 
\usepackage{multirow}
\usepackage{newtxtext}                                        
\usepackage{booktabs}
\usepackage[dvipsnames]{xcolor} 
\usepackage{hyperref}
\usepackage{float}
\usepackage[normalem]{ulem}
\usepackage[textwidth=1.5cm,textsize=footnotesize]{todonotes}
\hypersetup{colorlinks, 
    linkcolor={blue!75!black!80!yellow},
    citecolor={blue!75!black!80!yellow}, 
    urlcolor={blue!75!black!80!yellow}
    }
\usepackage[cmintegrals]{newtxmath}

\makeatletter
\renewcommand\@make@capt@title[2]{%
    \@ifx@empty\float@link{\@firstofone}{\expandafter\href\expandafter{\float@link}}%
    \sffamily{\textbf{#1}}\@caption@fignum@sep#2
}%

\makeatother

\newcommand{\HarvardSEAS}{John A. Paulson School of Engineering and Applied Sciences, Harvard University, Cambridge, MA 02138, USA}
\newcommand{\MaxPlanckDresden}{Max Planck Institute for Chemical Physics of Solids, 01187 Dresden, Germany}
\newcommand{\MITDMSE}{Department of Materials Science and Engineering, Massachusetts Institute of Technology, Cambridge, MA 02139, USA}

\begin{document} 

\author{Christina A. C. Garcia}\affiliation{\HarvardSEAS}
\author{Dennis M. Nenno}\affiliation{\HarvardSEAS}\affiliation{\MaxPlanckDresden}
\author{Georgios Varnavides}\affiliation{\HarvardSEAS}\affiliation{\MITDMSE}
\author{Prineha Narang}\email{prineha@seas.harvard.edu}\affiliation{\HarvardSEAS}

\title{Anisotropic phonon-mediated electronic transport in chiral Weyl semimetals}

\date{\today}

\begin{abstract}
\noindent Discovery and observations of exotic, quantized optical and electrical responses have sparked renewed interest in nonmagnetic chiral crystals.
Within this class of materials, six group V transition metal ditetrelides, that is, XY$_2$ (X\,=\,V,\,Nb,\,Ta and Y\,=\,Si,\,Ge), host composite Weyl nodes on high-symmetry lines, with Kramers-Weyl fermions at time-reversal invariant momenta.
In addition, at least two of these materials, NbGe$_2$ and NbSi$_2$, exhibit superconducting transitions at low temperatures.
The interplay of strong electron-phonon interaction and complex Fermi surface topology present an opportunity to study both superconductivity and hydrodynamic electron transport in these systems.
Towards this broader question, we present an \emph{ab initio} theoretical study of the electronic transport and electron-phonon scattering in this family of materials, with a particular focus on NbGe$_2$ vs. NbSi$_2$, and the other group V ditetrelides.
We shed light on the microscopic origin of NbGe$_2$'s large and anisotropic room temperature resistivity and contextualize its strong electron-phonon scattering with a presentation of other relevant scattering lifetimes, both momentum-relaxing and momentum-conserving.
Our work explores the intriguing possibility of observing hydrodynamic electron transport in these chiral Weyl semimetals.

\end{abstract}

\maketitle

Structurally chiral crystals have gained renewed attention, as they can realize Kramers-Weyl fermions at high-symmetry points of the band structure~\cite{Chang2018,manes2012existence,Tsirkin2017}.
Crystals that lack inversion, mirror or rotoinversion symmetries are intrinsically chiral, and it has been theoretically suggested that in nonmagnetic compounds which belong to chiral space groups, all crossing points at time-reversal invariant momenta (TRIM) in the electronic band structure carry a topological charge~\cite{manes2012existence,Chang2018}.
In systems where these topological bands dominate the electrical and optical response, a number of exciting phenomena have been proposed, such as nonlocal and nonreciprocal electron transport~\cite{Chang2018,narang2020topology} and a large quantized circular photogalvanic effect at room temperature~\cite{rees2019quantized}.

In this work, we investigate electronic transport in all transition metal ditetrelides, that is, XY$_2$ (X\,=\,V,\,Nb,\,Ta and Y\,=\,Si,\,Ge), known to crystallize in the non-symmorphic hexagonal chiral space group $P6_222$ (no. 180) or $P6_422$ (no. 181), with a particular focus on NbSi$_2$ and NbGe$_2$. 
Transition metal disilicides (Y\,=\,Si) have a rich literature primarily due to their low resistivity and possible application as silicon contacts~\cite{Gottlieb1991,hirano1990electrical,lasjaunias1993low}.
Their thermal and electrical properties have been studied in technological contexts~\cite{LABORDE2003415,wang2019insight,Gottlieb1991,lasjaunias1993low,BALKASHIN1996293,onuki2014chiral}, and their structural stability and electronic properties ranging from the superconducting transition and up to room temperature have also been investigated~\cite{wang2019insight,BALKASHIN1996293,antonov1996electronic}.
The experimental focus in group V digermanides (Y\,=\,Ge) has been driven by the low-temperature superconductivity exhibited by thin films and single crystals~\cite{Remeika1978}. 
They have been studied as potential high-temperature structural materials, in particular NbGe$_2$ with its type-I to type-II superconducting transition around 2~K and extremely high residual-resistance ratio in clean samples~\cite{lv2020type,Remeika1978}.
The combination of topological band character together with low-temperature superconductivity presents this whole family of materials as an intriguing  playground for fundamental questions in the strength of electron-phonon coupling in condensed matter.

Our study explores the anisotropic electronic transport properties of group V transition metal disilicides and digermanides through first principles calculations, incorporating the interactions between electronic and phononic states. 
In particular, we compute the intrinsic resistivities of these materials as functions of temperature, as they arise from electron-phonon interactions, and explore their respective possibilities for observing a hydrodynamic transport regime. 
We focus on the near order of magnitude higher resistivity in NbGe$_2$ compared to other group V ditetrelides at high temperatures. 
We trace this difference back to the pronounced scattering at the Fermi surface, which is energetically favored for certain stoichiometry. 
The strong electron-phonon interaction in NbGe$_2$ -- the origin of high room-temperature resistivity -- serves as a motivation to study the possibility of observing hydrodynamic transport induced by extremely short-lived, phonon-mediated electron-electron scattering events in this material. 
We predict the hydrodynamic electron flow in microscale devices of NbGe$_2$ of comparable size to be even more pronounced than in the recently reported Poiseuille flow observed in WTe$_2$~\cite{wte2-hydro} wires.

\begin{figure}[t]
    \centering
    \includegraphics[width=0.5\textwidth]{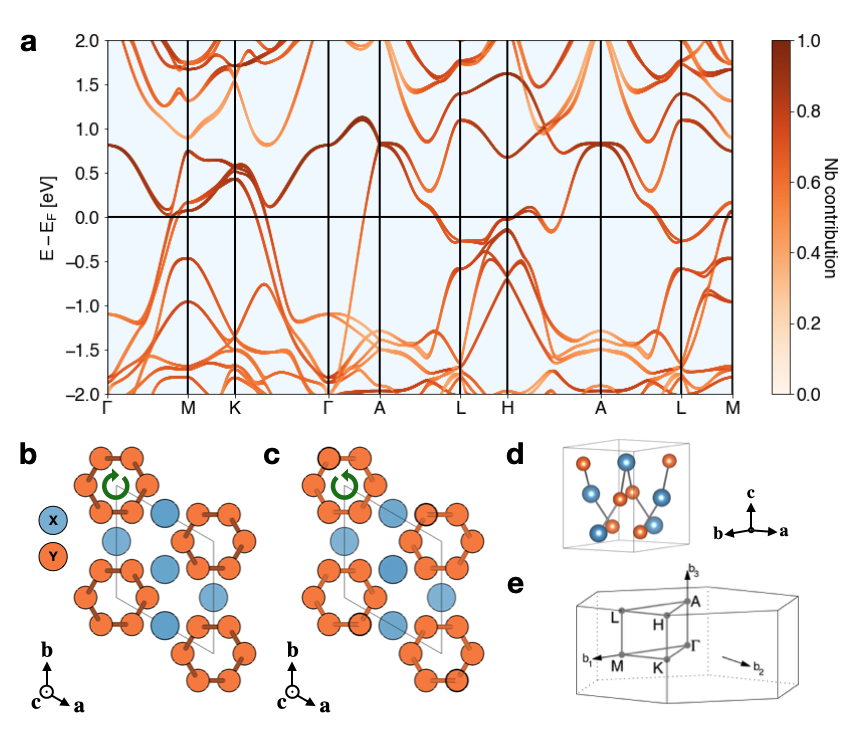}
    \caption{(a) Electronic band structure of NbGe$_2$ in space groups 180 and 181 color-coded by Nb-contribution to the bands. 
    The unit cell of NbGe$_2$ in space group 180 (b) and spacegroup 181 (c) is displayed viewed along the $c$-axis and from the side (d). 
    (e) Brillouin zone of the hexagonal unit cell with high-symmetry paths (from Ref.~\citenum{SETYAWAN2010299}).}
    \label{fig:1}
\end{figure}

\begin{figure*}[t]
    \centering
    \includegraphics[width=\textwidth]{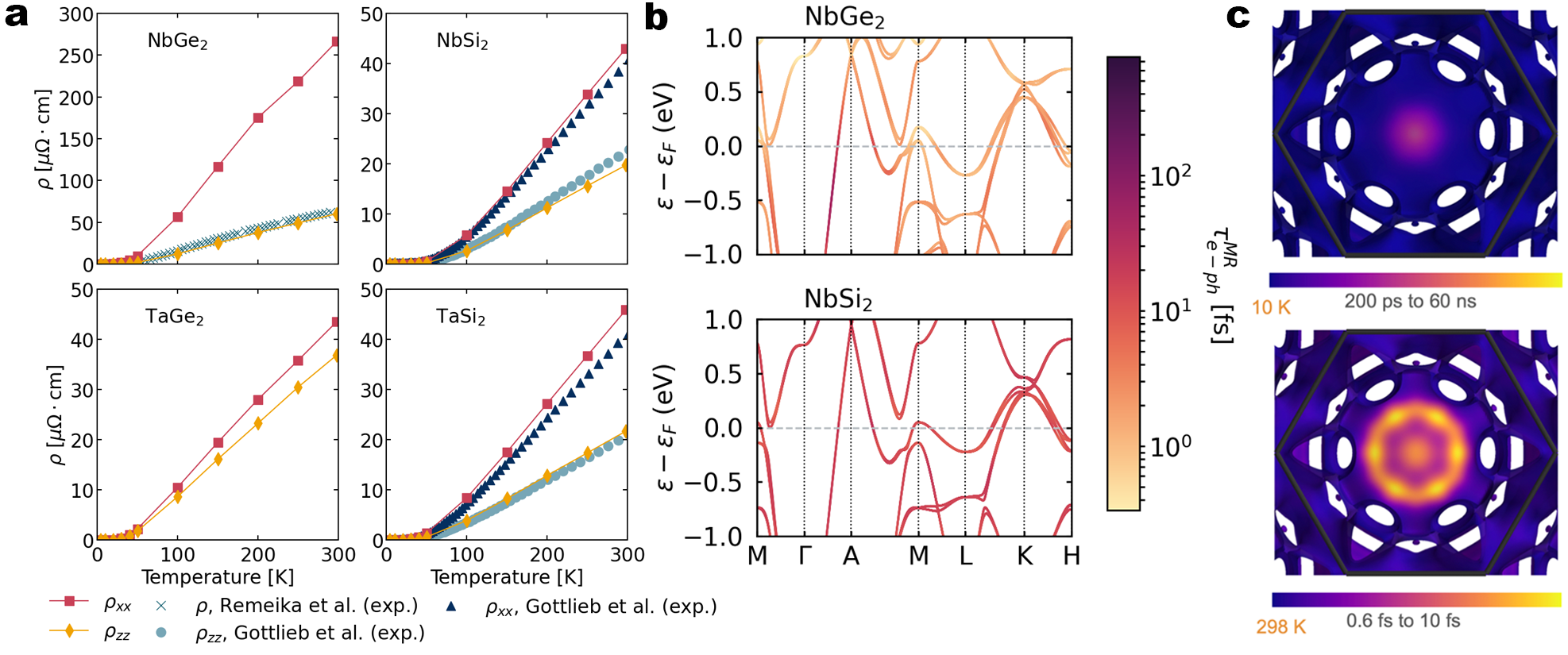}
    \caption{(a) Resistivity vs. temperature for NbGe$_2$, NbSi$_2$, TaGe$_2$, and TaSi$_2$. 
    Experimental data by Remeika \emph{et al.}~\cite{Remeika1978} and Gottlieb \emph{et al.}~\cite{Gottlieb1991} are overlaid for comparison. 
    (b) Computed room temperature momentum-relaxing electron-phonon lifetimes plotted on the electronic band structures of NbGe$_2$ and NbSi$_2$. 
    While there are small variations in these lifetimes across the band structures for all six materials, the lifetimes of NbGe$_2$ are roughly an order of magnitude shorter than those of NbSi$_2$ and the other five materials (see Supplemental Material), indicating stronger momentum-relaxing electron-phonon (e-ph) scattering. 
    This stronger e-ph scattering can be explained by one to two orders of magnitude difference in the e-ph coupling. 
    (c) The electron-phonon lifetimes plotted on the Fermi surface of NbGe$_2$ for 10~K and 298~K shown on the (0001)-surface. The hexagon marks the boundary of the first Brillouin zone.}
    \label{fig:scattering}
\end{figure*}

All six transition metal ditetrelides show stable phases in both the nonsymmorphic Sohnke space group $P6_222$ (no. 180) and its energy-degenerate enantiomorphic partner space group $P6_422$ (no. 181) and can be classified as enforced semimetals with Fermi degeneracy~\cite{bradlyn2017topological,vergniory2019complete} 
The band structure for NbGe$_2$, shown in Fig.~\ref{fig:1}(a), coincides for both left- and right-handed chiralities of the unit cell (cf.~Fig.~\ref{fig:1}(b,c)). 
All six compounds contain topologically nontrivial features in their electronic structure, owing to crystalline symmetries and strong spin-orbit coupling~\cite{Chang2018}. 
Kramers-Weyl nodes~\cite{Zhang2018} are found at TRIM, i.e. the $\Gamma$, $M$, $A$ and $L$ points of the hexagonal Brillouin zone shown in Fig.~\ref{fig:1}(e). 
These degenerate crossings are enforced by the crystal symmetries, and the crossings closest to the Fermi surface appear at the $M$ and $H$ points in NbGe$_2$, where the dispersion becomes flat. 
Along the sixfold rotation symmetry axis $\Gamma$-$A$, band crossings with four-fold degeneracies occur, which are protected due to their opposite eigenvalues of the sixfold screw rotations $C_{6,2}$~\cite{Tsirkin2017,manes2012existence}. 
However, in all six compounds, these points lie more than 1\,eV below the Fermi level, and only one doubly degenerate band crosses the Fermi surface along this line, which has a strong impact on the resistivity along $\hat{z}$. 
The band structure close to the Fermi level shows two pairs of bands split by spin-orbit coupling along the $\Gamma$-$M$ axis, which lead to two doubly degenerate Kramers-Weyl points (chirality $|\chi|=1$) on the TRIM. 
As observed in structurally-chiral PdGa, the chiral charge at each of these crossings flips sign between the enantiomorphic partners due to mirror symmetry~\cite{schroter2019observation}. 
Consequently, it is possible to distinguish the two enantiomers by their surface states. 
Due to bulk-surface correspondence, Fermi arcs form between Weyl points of opposite chirality. 

In NbGe$_2$, a hole pocket emerges from unoccupied states at $\sim$5 meV above the Fermi level. 
The density of states, and thus the scattering state space at higher temperatures, increases drastically close to the bottom of this band and the Kramers-Weyl nodes in the vicinity of the Fermi level. 
While in the topological semimetal RhSi there exists a large energy window in the vicinity of the Fermi level in which topological degeneracies dominate and lead to a number of quantized responses, in NbGe$_2$, we do not expect Weyl points to dictate the electronic response~\cite{PhysRevLett.119.206402,rees2019quantized}.

We compute the DC electrical resistivity tensor $\rho$ from first principles using a linearized Boltzmann transport equation within a full-band relaxation time approximation incorporating state-resolved momentum-relaxing scattering lifetimes $\tau_{\text{e-ph}}^{\text{MR}}(\textbf{k},n)$ and velocities $v_{n\textbf{k}}$ (see Supplemental Material for more details and Refs.~\citenum{ACSNanoBrown2016,Garcia2020,Coulter2018,Ciccarino2018a} therein),
\begin{equation}
    \bar{\rho}^{-1}=\int_{\text{BZ}}\frac{e^2d\textbf{k}}{(2\pi)^3}\sum_{n}\frac{\partial f_{n\textbf{k}}}{\partial \varepsilon_{n\textbf{k}}}
    (v_{n\textbf{k}}\otimes v_{n\textbf{k}})\,\tau_{\text{e-ph}}^{\text{MR}}(\textbf{k},n)\,,
    \label{eq:sigma0_maintext}
\end{equation}
where we integrate over all momenta $\textbf{k}$ in the full Brillouin zone and sum over all electronic bands $n$, with $\varepsilon_{n\textbf{k}}$ and $f_{n\textbf{k}}$ being the electronic energies and occupations.

\begin{figure}[b]
    \centering
    \includegraphics[width=0.49\textwidth]{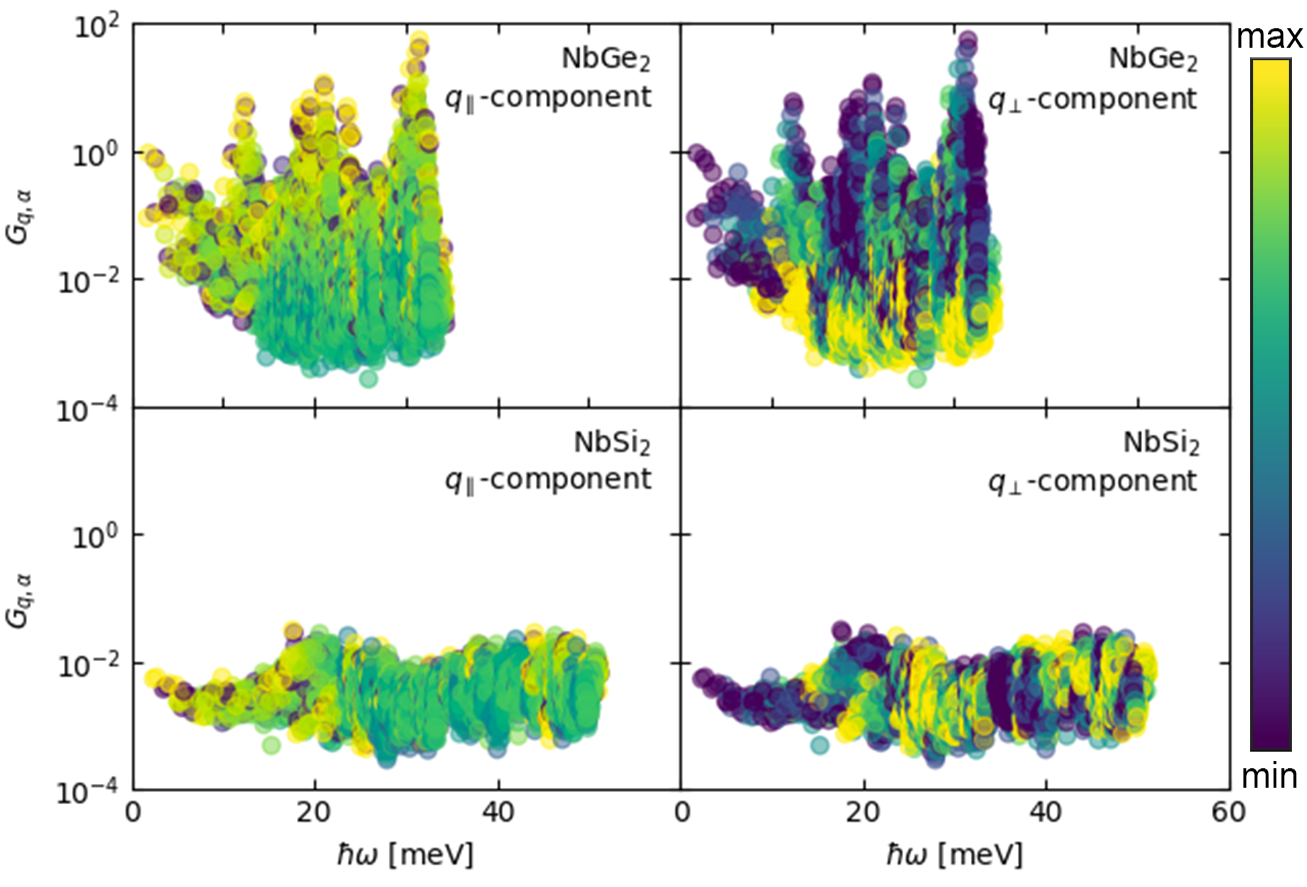}
    \caption{Fermi-surface averaged electron-phonon coupling strength $G$ for various phonon modes $\alpha$ and momenta $q$ (from Eq.~\eqref{eq:geph} in the Supplemental Material). For both NbGe$_2$ (upper panels) and NbSi$_2$ (lower panels), the color encodes the magnitude of the inplane, $q_{\parallel}$ (left), and out-of-plane component, $q_\perp$ (right), of the phonon momentum, ranging from 0 (blue) to maximum (yellow).}
    \label{fig:Geph}
\end{figure}

The results for varying temperature in NbGe$_2$, NbSi$_2$, TaGe$_2$, and TaSi$_2$ are shown in Fig.~\ref{fig:scattering}(a), with available experimental data for NbGe$_2$, NbSi$_2$, and TaSi$_2$ overlaid~\cite{Remeika1978,Gottlieb1991}. 
Similarly, for the vanadium ditetrelides we calculate $\rho_{xx}=72.3$\,$\mu\Omega$\,cm and $\rho_{zz}=38.4$\,$\mu\Omega$\,cm for VGe$_2$ and $\rho_{xx}=58.1$\,$\mu\Omega$\,cm and $\rho_{zz}=23.8$\,$\mu\Omega$\,cm for VSi$_2$ at room temperature.
Since we only account for the contribution of momentum-relaxing electron-phonon scattering, we expect to underestimate the low-temperature resistivity unless the crystal is very clean, i.e. the scattering of electrons with impurities is negligible. 
This likely explains the mild underestimation seen for $\rho_{zz}$ in NbGe$_2$ and NbSi$_2$ at lower temperatures. 
A slight overestimation of the resistivity, as seen for $\rho_{xx}$ in NbSi$_2$ and TaSi$_2$ and $\rho_{zz}$ in TaSi$_2$, can be explained by a misaligned Fermi level compared to the measured sample or slight differences in the lattice constants between our theoretical calculations and experiment.
However, the overall excellent agreement with experiment indicates that the level of theory used is appropriate to capture the high-temperature behavior of the resistivity.

In comparing the high-temperature resistivities of these materials, we see that NbGe$_2$ stands out as having significantly larger resistivities along both $a$ ($xx$) and $c$ ($zz$) crystallographic directions, with $\rho_{xx}$ and $\rho_{zz}$ in NbGe$_2$ at room temperature being 3.7$\times$ and 1.6$\times$ those in VGe$_2$, which has the largest resistivities among the remaining five compounds.
Only the in-plane resistivity has been accessed in experiments so far.
NbGe$_2$ also has the largest anisotropy in resistivity between the $a$ and $c$ directions, with $\rho_{xx}/\rho_{zz}\approx 4.4$ at room temperature according to our predictions. 
Among the other group V transition metal ditetrelides, this ratio is 2.2, 1.2, 2.1, 1.9, and 2.4 for NbSi$_2$, TaGe$_2$, TaSi$_2$, VGe$_2$, and VSi$_2$, respectively. 
These results are surprising given that all six compounds considered here are isostructural and isoelectronic.

\begin{figure*}[t]
    \centering
    \includegraphics[width=\textwidth]{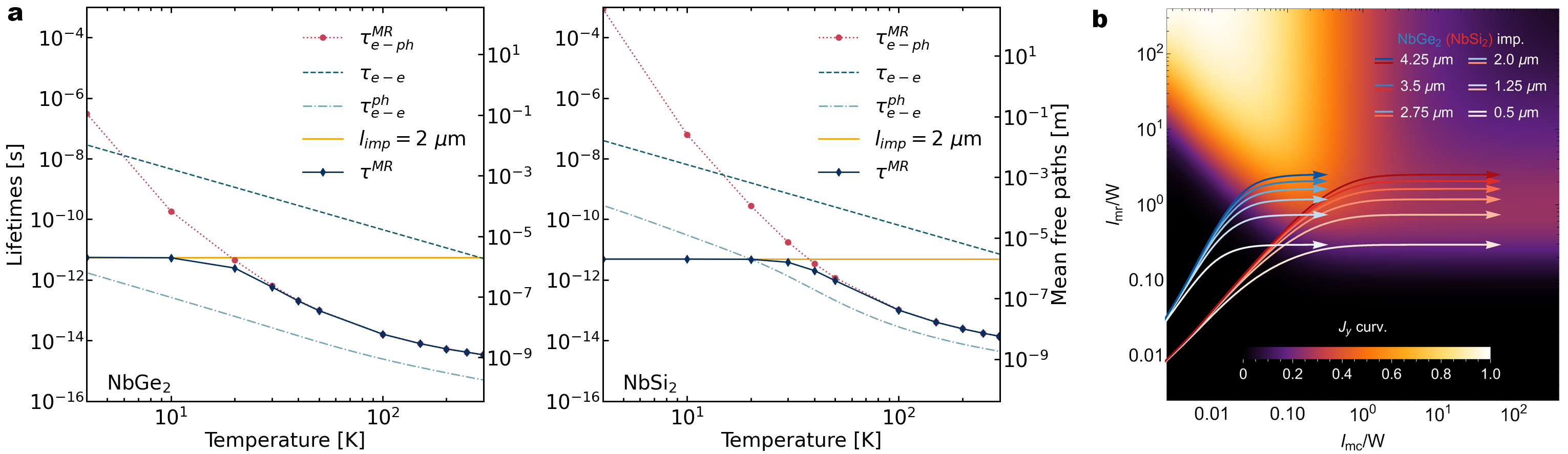}
    \caption{(a) Predicted temperature-dependent lifetimes averaged over the Fermi surface for different scattering processes in NbGe$_2$ and NbSi$_2$. 
    The momentum-relaxing electron-phonon scattering lifetime ($\tau_{\text{e-ph}}^{\text{MR}}$) is shorter than the momentum-conserving Coulomb-mediated electron-electron scattering lifetime ($\tau_{\text{e-e}}$) until low temperatures in both materials. 
    The phonon-mediated electron-electron scattering lifetime ($\tau_{\text{e-e}}^{\text{ph}}$) is shorter than these and is the dominant scattering process at all temperatures. 
    At low temperatures, small-angle scattering dominates and suppresses momentum-relaxing electron-phonon scattering, as evident in comparing the electron-phonon scattering lifetime ($\tau_{\text{e-ph}}$) to $\tau_{\text{e-ph}}^{\text{MR}}$.
    (b) Prediction of hydrodynamic flow regimes in NbGe$_2$ (blue lines) and NbSi$_2$ (red lines) for different impurity mean-free paths. 
    The colorscale denotes the curvature of the current profile in a channel of width $W$ for different momentum-relaxing ($l_{\text{mr}}$) and momentum-conserving ($l_{\text{mc}}$) mean free paths. 
    The superimposed trajectories, obtained from our \textit{ab initio} calculations, in NbGe$_2$ (blue) and NbSi$_2$ (red), for a channel width of 1~$\mu$m, run from 298~K to 4~K.}
    \label{fig:lifetimes}
\end{figure*}

To understand why the electrical resistivity of NbGe$_2$ at room temperature is so much larger and more anisotropic than the other group V transition metal ditetrelides, we analyze the main contributor to the resistivity, the momentum-relaxing electron-phonon scattering lifetime ($\tau_{\text{e-ph}}^{\text{MR}}$) (see Supplemental Material). 
In Fig.~\ref{fig:scattering}(b), we present momentum- and energy-resolved $\tau_{\text{e-ph}}^{\text{MR}}$ along high-symmetry directions for NbGe$_2$ and NbSi$_2$ at room temperature. 
While there is noticeable variation in this lifetime across each band structure, it is apparent in comparing the two materials that this lifetime is roughly an order of magnitude shorter overall, i.e. across energy and momentum, in NbGe$_2$ compared to NbSi$_2$.
This indicates that NbGe$_2$ has stronger momentum-relaxing electron-phonon scattering and helps to explain its comparatively larger resistivity, which is inversely proportional on the $\tau_{\text{e-ph}}^{\text{MR}}$, increasing with increased momentum-relaxing scattering.

In addition, we note that for the band intersecting the Fermi level along the $k$-path $\Gamma$-$A$ in NbGe$_2$, $\tau_{\text{e-ph}}^{\text{MR}}$ is not as short as the same lifetime at other points that also cross the Fermi level; instead, it is comparable to the lifetimes calculated for NbSi$_2$ and the other transition metal ditetrelides (see Supplemental Material). 
As transitions along $\Gamma$-$A$ contribute primarily to $\rho_{zz}$ and transitions along $\Gamma$-$M$ contribute primarily to $\rho_{xx}$, the significant difference in momentum-relaxing scattering along these directions accounts for part of the large anisotropy we predict for NbGe$_2$. 
Separately, the increased scattering along $M$-$L$ and $K$-$H$, which also contributes to $\rho_{zz}$, results in $\rho_{zz}$ still being much larger in NbGe$_2$ than in the other five materials, despite the lesser scattering along $\Gamma$-$M$. 
These observations regarding the momentum-relaxing electron-phonon scattering lifetimes help to explain the resistivity results shown in Fig.~\ref{fig:scattering}(a). 

We note, however, that the electronic band structures shown in Fig.~\ref{fig:scattering} do not provide insight into momentum-relaxing scattering across the whole Fermi surface and thus must be regarded as a glimpse rather than a comprehensive picture.
By contrast, Fig.~\ref{fig:scattering}(c) plots $\tau_{\text{e-ph}}^{\text{MR}}$ on the Fermi surface of NbGe$_2$ at 10~K and 298~K.
At both temperatures, only part of the Fermi surface, with velocities along the $z$-axis, centered around the $\Gamma$ point, shows the longest lifetimes, while other regions show lifetimes as low as 0.6\,fs.

To better understand the order of magnitude difference in $\tau_{\text{e-ph}}^{\text{MR}}$ of NbGe$_2$ versus those of the other five group V transition metal ditetrelides, we compute the Fermi-surface averaged electron-phonon interaction strength $G$ (see Supplemental Material) for different phonon wavevectors $q$ and modes $\alpha$.
We present these for NbGe$_2$ and NbSi$_2$ in Fig.~\ref{fig:Geph}.
The result provides microscopic insight into the anisotropy: while the modes that couple strongly, and thus lead to a reduced lifetime and increased resistivity, cluster depending on their phonon momentum direction in NbGe$_2$, the coupling in NbSi$_2$ appears to be more isotropic.
In particular, the phonons that can transfer momentum in the crystallographic $z$-direction in NbGe$_2$ are almost exclusively weakly coupled, while phonons that can transfer electron momentum in-plane exhibit stronger couplings .
A more detailed exposition of the electron-phonon coupling strength and its mode- and angular-dependence in NbGe$_2$ and NbSi$_2$ is presented in the Supplemental Material.

In predicting the transport behavior of the group V transition metal ditetrelides using momentum-relaxing electron-phonon scattering rates, it is also useful to understand this type of scattering in the context of other scattering mechanisms present in the material. 
To this end, we compute the screened Coulomb electron-electron scattering lifetime ($\tau_{\text{e-e}}$), the phonon-mediated electron-electron scattering lifetime ($\tau_{\text{e-e}}^{\text{ph}}$), and the electron-phonon scattering lifetime ($\tau_{\text{e-ph}}$), different from the momentum-relaxing electron-phonon scattering lifetime in that small-angle scattering events are no longer weighted out (see Supplemental Material).
We plot these scattering lifetimes vs. temperature for NbGe$_2$ and NbSi$_2$ in Fig.~\ref{fig:lifetimes}(a). 
These scattering processes can be further classified based on whether they conserve or relax the total momentum of the electronic system.
Seen in this light, the screened Coulomb and phonon-mediated electron-electron interactions are momentum-conserving, while both the electron-phonon and electron-impurity interactions are momentum-relaxing.
While the timescale for direct Coulomb electron-electron scattering processes in both materials is comparable for all temperatures, the stronger electron-phonon coupling in NbGe$_2$ drastically reduces the momentum-conserving electron-electron lifetime mediated by a virtual phonon. 

Noting that momentum-conserving scattering in NbGe$_2$ is the fastest scattering mechanism reported here for all temperatures, we investigate the potential for hydrodynamic transport in NbGe$_2$ and NbSi$_2$.
Phonon-mediated anisotropic hydrodynamic flow has recently been the focus of both theoretical investigations~\cite{Levchenko2020,Varnavides2020,Bradlyn2020,Lucas2020}, and direct spatially-resolved imaging in ultra-pure samples of WTe$_2$ for the first time~\cite{wte2-hydro}.
To this end, it is instructive to relate the Fermi-surface averaged scattering lifetimes to a momentum-conserving ($l_{\text{mc}}$) and a momentum-relaxing ($l_{\text{mr}}$) mean-free path, using the respective Fermi velocities.
Relative to the geometry's length-scale ($W$), these mean free paths then specify the electron flow regime we are sampling as a function of temperature~\cite{molenkamp-hydro,wte2-hydro}.
Figure~\ref{fig:lifetimes}(b) plots the curvature of the current density for a large range of these two non-dimensional parameters~\cite{wte2-hydro}, illustrating that in order to observe electron hydrodynamic flow we require $l_{\text{mc}}\ll W < l_{\text{mr}}$.
Superimposing the temperature-dependent mean-free paths, $l_{\text{mc}}(T)$ and $l_{\text{mr}}(T)$, for various empirical values of impurity length-scales~\cite{Gottlieb1991}, we see that in relatively clean samples (impurity mean free paths larger than 2~$\mu$m), both NbGe$_2$ and NbSi$_2$ develop curved current profiles, suggesting non-diffusive behavior for micron-scale geometries. 
NbGe$_2$ shows a much larger current density curvature, with a peak curvature of J$_y$ curv. $\sim$0.525 at $\sim$7.5~K, while the non-diffusive behavior in NbSi$_2$ persists at higher temperatures, with a peak curvature of J$_y$ curv. $\sim$0.25 at $\sim$35~K. 
This value exceeds the measured and predicted result in WTe$_2$ and thus makes NbGe$_2$ a desirable platform for further exploration of phonon-mediated hydrodynamic transport~\cite{wte2-hydro}.

In conclusion, we have presented \emph{ab initio} transport calculations for all six transition metal ditetrelides that crystallize in chiral hexagonal unit cells.
We find excellent agreement with experiments for the subset of these materials whose electrical properties have been characterized, and at high temperatures we predict an emphasized anisotropy in the resistivity of NbGe$_2$ not observed by experiments so far. 
Importantly, we find that the overall comparatively high resistivities in NbGe$_2$ originate from strong electron-phonon interaction, which also leads to a higher superconducting transition temperature in this compound, as observed, for example, by Ref.~\cite{knoedler1979superconductivity}. 
Looking ahead, we observe that in moderately low temperatures above the superconducting transition, this strong electron-phonon coupling leads to pronounced momentum-conserving electron-electron scattering due to the exchange of virtual phonons, making NbGe$_2$ an ideal candidate to observe phonon-mediated hydrodynamic flow. We anticipate future work in this class of materials on the interplay and competition between hydrodynamic transport and superconductivity.

\bibliography{references}

\section*{Acknowledgements}
We are grateful for discussions with Dr. Yaxian Wang at Harvard University and Prof. Dr. Claudia Felser at the Max Planck Institute for Chemical Physics of Solids. This work is supported by the STC Center for Integrated Quantum Materials, NSF Grant No. DMR-1231319 for development of computational methods for topological materials. 
G.V. and P.N. acknowledge support from the Army Research Office MURI (Ab-Initio Solid-State Quantum Materials) Grant No. W911NF-18-1-0431 that supported development of computational methods to describe microscopic, temperature-dependent dynamics in low-dimensional materials. G.V. acknowledges support from the Office of Naval Research grant on High-Tc Superconductivity at Oxide-Chalcogenide Interfaces (Grant Number N00014-18-1-2691) that supported theoretical and computational methods for phonon-mediated interactions.
This work used resources of the National Energy Research Scientific Computing Center, a DOE Office of Science User Facility, as well as resources at the Research Computing Group at Harvard University. 
Additional calculations were performed using resources from the Department of Defense High Performance Computing Modernization Program through the Army Research Office MURI grant on Ab-Initio Solid-State Quantum Materials: Design, Production, and Characterization at the Atomic Scale (18057522). 
C.A.C.G. was supported by the NSF Graduate Research Fellowship Program under Grant No. DGE-1745303. P.N. is a Moore Inventor Fellow and gratefully acknowledges support through Grant No. GBMF8048 from the Gordon and Betty Moore Foundation.

\newpage
\onecolumngrid
\appendix
\section*{Supplemental Material}

\subsection*{Scattering lifetimes}

We compute the lifetimes associated with four types of microscopic scattering events in NbGe$_2$ and NbSi$_2$: electron-phonon scattering ($\tau_{\text{e-ph}}$), momentum-relaxing electron-phonon scattering ($\tau^{\text{MR}}_{\text{e-ph}}$), electron-electron scattering mediated by the Coulomb interaction ($\tau_{\text{e-e}}$), and phonon-mediated electron-electron scattering ($\tau^{\text{ph}}_{\text{e-e}}$). 
Our methodology, for the most part, follows that presented in previous works~\cite{ACSNanoBrown2016,Garcia2020,Coulter2018,Ciccarino2018a}

We calculate the temperature-dependent, state-resolved electron-phonon scattering rate using Fermi's golden rule, as presented in Refs.~\citenum{ACSNanoBrown2016,Garcia2020,Coulter2018,Ciccarino2018a}:
\begin{equation}
    \frac{1}{\tau_{\text{e-ph}}(\textbf{k},n)}
    =\frac{2\pi}{\hbar}\int_{\text{BZ}}\frac{\Omega\, d\textbf{k}'}{(2\pi)^3}\sum_{n'\alpha\pm}
    \delta(\varepsilon_{n'\textbf{k}'}-\varepsilon_{n\textbf{k}}\mp\hbar\omega_{\alpha\textbf{k}'-\textbf{k}})
    \times\left[n_{\alpha\textbf{k}'-\textbf{k}}+\frac{1}{2}\mp\left(\frac{1}{2}-f_{n'\textbf{k}'}\right)\right]
    \left|g_{n'\textbf{k}',n\textbf{k}}^{\alpha\textbf{k}'-\textbf{k}}\right|^2
    \label{eq:tau_e-ph}
\end{equation}

\noindent where $\Omega$ is the unit cell volume, $\varepsilon_{n\textbf{k}}$ and $f_{n\textbf{k}}=f(\varepsilon_{n\textbf{k}},T)$ are energies and Fermi occupations for an electron in band $n$ at wave vector \textbf{k}, $\hbar\omega_{\alpha\textbf{q}}$ and $n_{\alpha\textbf{q}}=n(\hbar\omega_{\alpha\textbf{q}},T)$ are energies and Bose occupations for a phonon with polarization index $\alpha$ at wave vector $\mathbf{q}$, and $g_{n'\textbf{k}',n\textbf{k}}^{\alpha\textbf{q}}$ is the electron-phonon coupling matrix element:

\begin{equation}
    g_{n'\textbf{k}',n\textbf{k}}^{\alpha\textbf{q}}=
    \left(\frac{\hbar}{2m_0\omega_{\alpha\textbf{q}}}\right)^{1/2}
    \left<\psi_{n'\textbf{k}'}\right|\partial_{\alpha\textbf{q}}V\left|\psi_{n\textbf{k}}\right>\,.
\end{equation}

\begin{figure*}[t]
    \centering
    \includegraphics[width=\textwidth]{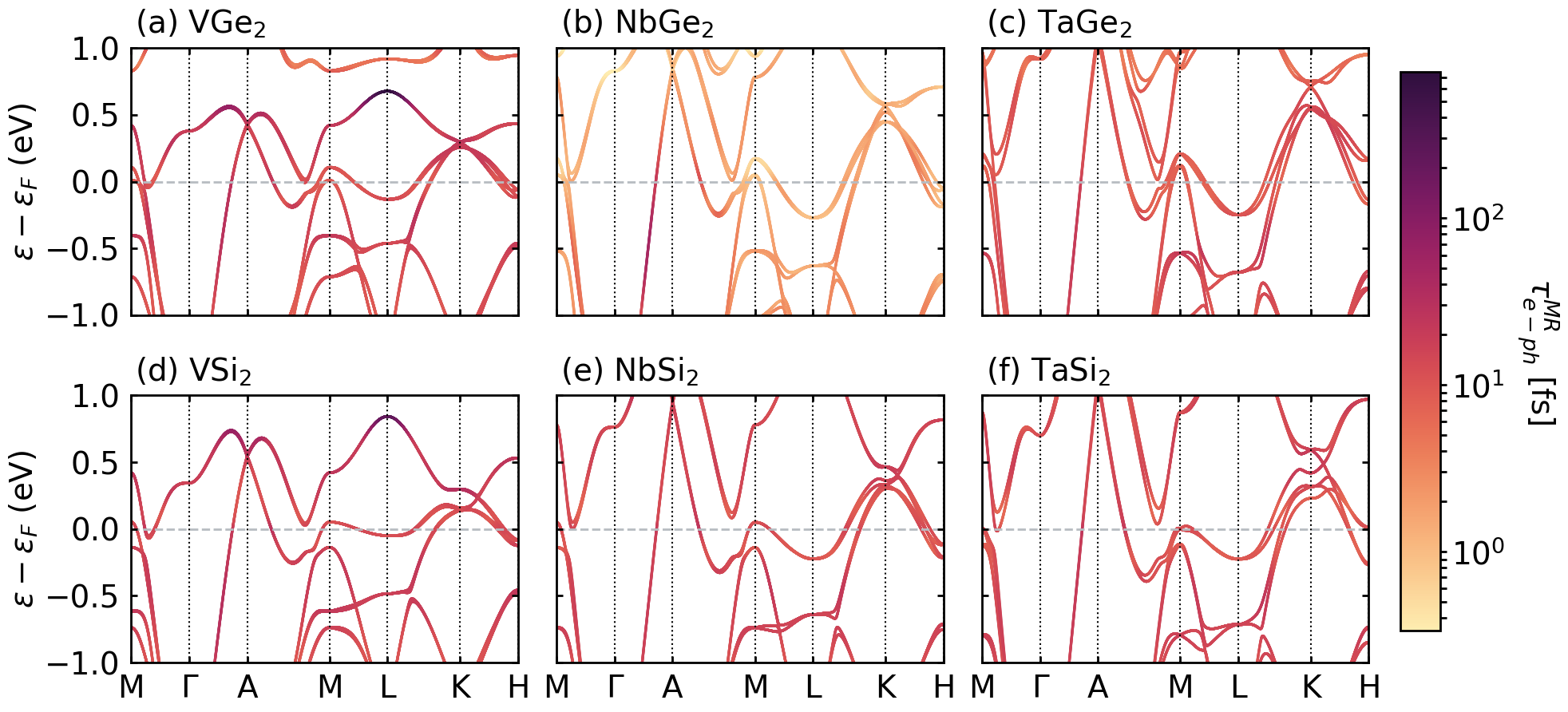}
    \caption{Computed room temperature momentum-relaxing electron-phonon lifetimes from Eq.~\ref{eq:tau_MR} plotted on the electronic band structures for all six transition metal ditetrelides that crystallize in space group 180 or 181.}
    \label{fig:tau_elph_all}
\end{figure*}

\noindent By momentum conservation, only processes where $\textbf{q}=\textbf{k}'-\textbf{k}$ are allowed, and this is already substituted into (\ref{eq:tau_e-ph}).
Phonon emission and absorption processes are accounted for in (\ref{eq:tau_e-ph}) with the sum over $\pm$.
The momentum relaxing (MR) scattering rate is identical to (\ref{eq:tau_e-ph}) except for an additional factor accounting for the change in momentum between final and initial states based on their relative scattering angle:

\begin{equation}
\begin{split}
    \frac{1}{\tau_{\text{e-ph}}^{\text{MR}}(\textbf{k},n)}
    =\frac{2\pi}{\hbar}\int_{\text{BZ}}\frac{\Omega\, d\textbf{k}'}{(2\pi)^3}\sum_{n'\alpha\pm}
    \delta(\varepsilon_{n'\textbf{k}'}-\varepsilon_{n\textbf{k}}\mp\hbar\omega_{\alpha\textbf{k}'-\textbf{k}})
    \times\left[n_{\alpha\textbf{k}'-\textbf{k}}+\frac{1}{2}\mp\left(\frac{1}{2}-f_{n'\textbf{k}'}\right)\right]
    \left|g_{n'\textbf{k}',n\textbf{k}}^{\alpha\textbf{k}'-\textbf{k}}\right|^2
     \left(1-\frac{v_{n\textbf{k}}\cdot v_{n'\textbf{k}'}}{\left|v_{n\textbf{k}}\right|\left|v_{n'\textbf{k}'}\right|}\right)\,,
\end{split} 
\label{eq:tau_MR}
\end{equation}

\noindent where $v_{n\textbf{k}}$ is the group velocity. The resulting lifetimes for all six transition metal ditetrelides discussed are presented in Fig.~\ref{fig:tau_elph_all}.

To put the momentum-relaxing scattering into perspective, we also calculate lifetimes for electron-electron scattering mediated by both the Coulomb interaction and virtual phonons.
The former is obtained by evaluating
\begin{equation}
\begin{split}
       \frac{1}{\tau_\mathrm{e-e}(\textbf{k},n)}=2\int_{\mathrm{BZ}}\frac{d\textbf{k}'}{(2\pi)^3}\sum_{n'}\sum_{\textbf{GG}'}\tilde{\rho}_{n'\textbf{k}',n\textbf{k}}(\textbf{G})
       \tilde{\rho}^*_{n'\textbf{k}',n\textbf{k}}(\textbf{G}')\times\frac{4\pi e^2}{|\textbf{k}'-\textbf{k}+\textbf{G}|^2}\mathrm{Im}[\epsilon^{-1}_{\textbf{GG}'}(\textbf{k}'-\textbf{k},\varepsilon_{n\textbf{k}}-\varepsilon_{n'\textbf{k}'})],
\end{split} 
\label{eq:tau_ee}
\end{equation}
where $\tilde{\rho}_{n'\textbf{k}',n\textbf{k}}(\textbf{G})$ is the plane wave expansion of the product density $\sum_{\sigma}u^{\sigma*}_{n'\textbf{k}'(\textbf{r})}u^{\sigma}_{n\textbf{k}}(\textbf{r})   $ of the Bloch functions with reciprocal lattice vectors \textbf{G}, and $\epsilon^{-1}_{\textbf{GG}'}(\textbf{k}'-\textbf{k},\varepsilon_{n\textbf{k}}-\varepsilon_{n'\textbf{k}'})$ is the microscopic dielectric function in a plane wave basis calculated within the random-phase approximation.
The electron-electron lifetime~\eqref{eq:tau_ee} is analytically extended to finite temperature values using a Fermi-liquid theory fit to the state-resolved result at zero temperature, i.e. by fitting,~\cite{ACSNanoBrown2016}
\begin{equation}
    \tau^{-1}_{\mathrm{e-e}}(\varepsilon,T)\approx \frac{D_e}{\hbar}[(\varepsilon-\varepsilon_{\mathrm{F}})^2+(\pi k_B T)^2]\,,
\end{equation}
where $\varepsilon_F$ is the Fermi energy.

\begin{figure*}[t]
    \centering
    \includegraphics[width=0.85\textwidth]{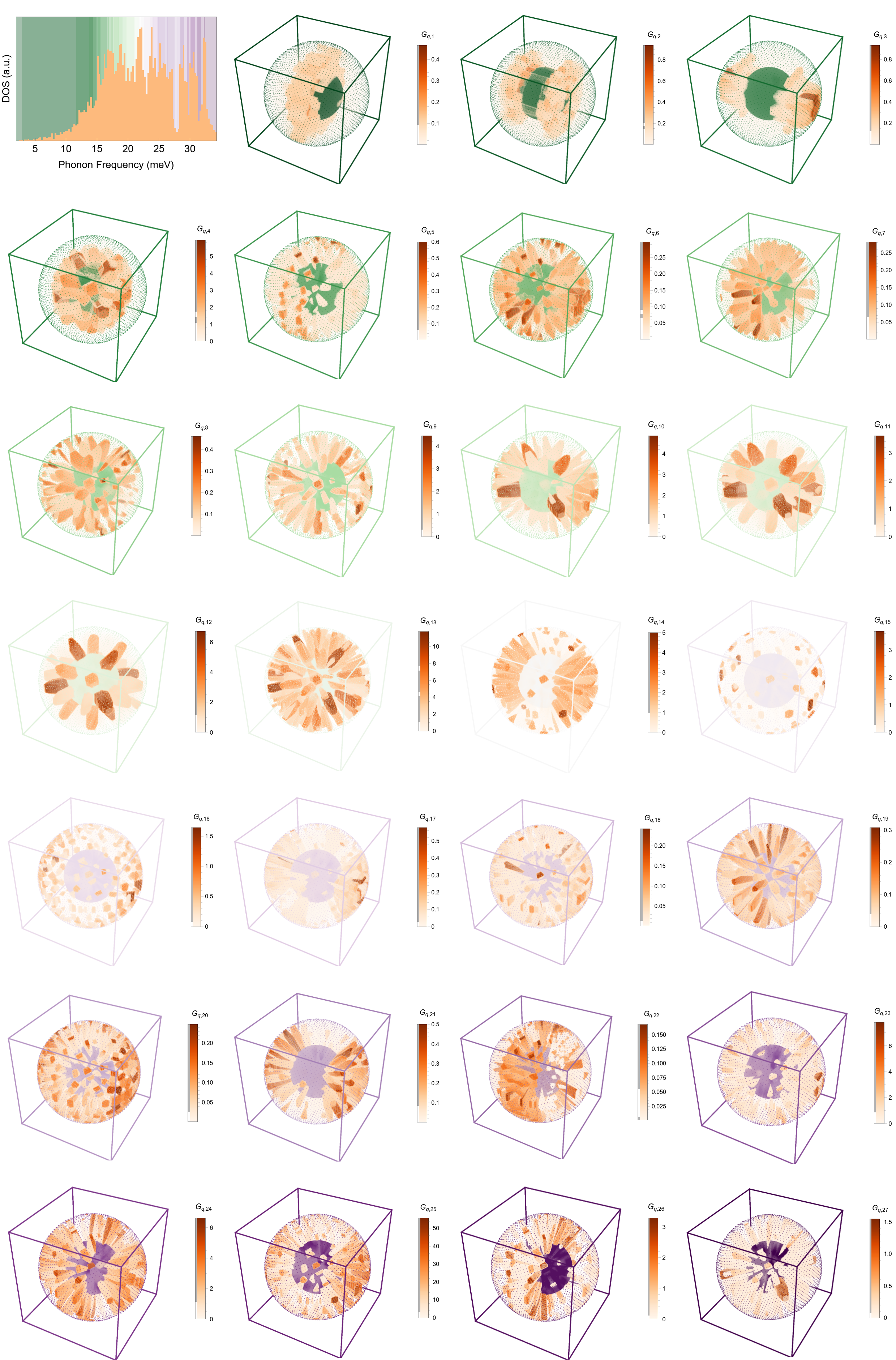}
    \caption{Phonon density of states (orange) of NbGe$_2$ (top left panel), together with mode-resolved Fermi-surface averaged electron-phonon coupling strength (the energy window for each mode is given by the color-coded background rectangles in the top left panel from low (green) to high (purple), also reflected in the box frame). 
    Subsequent panels show the angular-dependence in the phonon momentum $q$ of the Fermi-surface averaged electron-phonon coupling strength $G$ for various phonon modes $\alpha$, inside a spherical shell given by the mode's energy window. 
    The color-coding of the vectors represents the magnitude of $G$.
    Modes 10-12 exhibit strong in-plane anisotropy with strong electron-phonon couplings.}
    \label{fig:G_eph_NbGe2}
\end{figure*}

\begin{figure*}[t]
    \centering
    \includegraphics[width=0.85\textwidth]{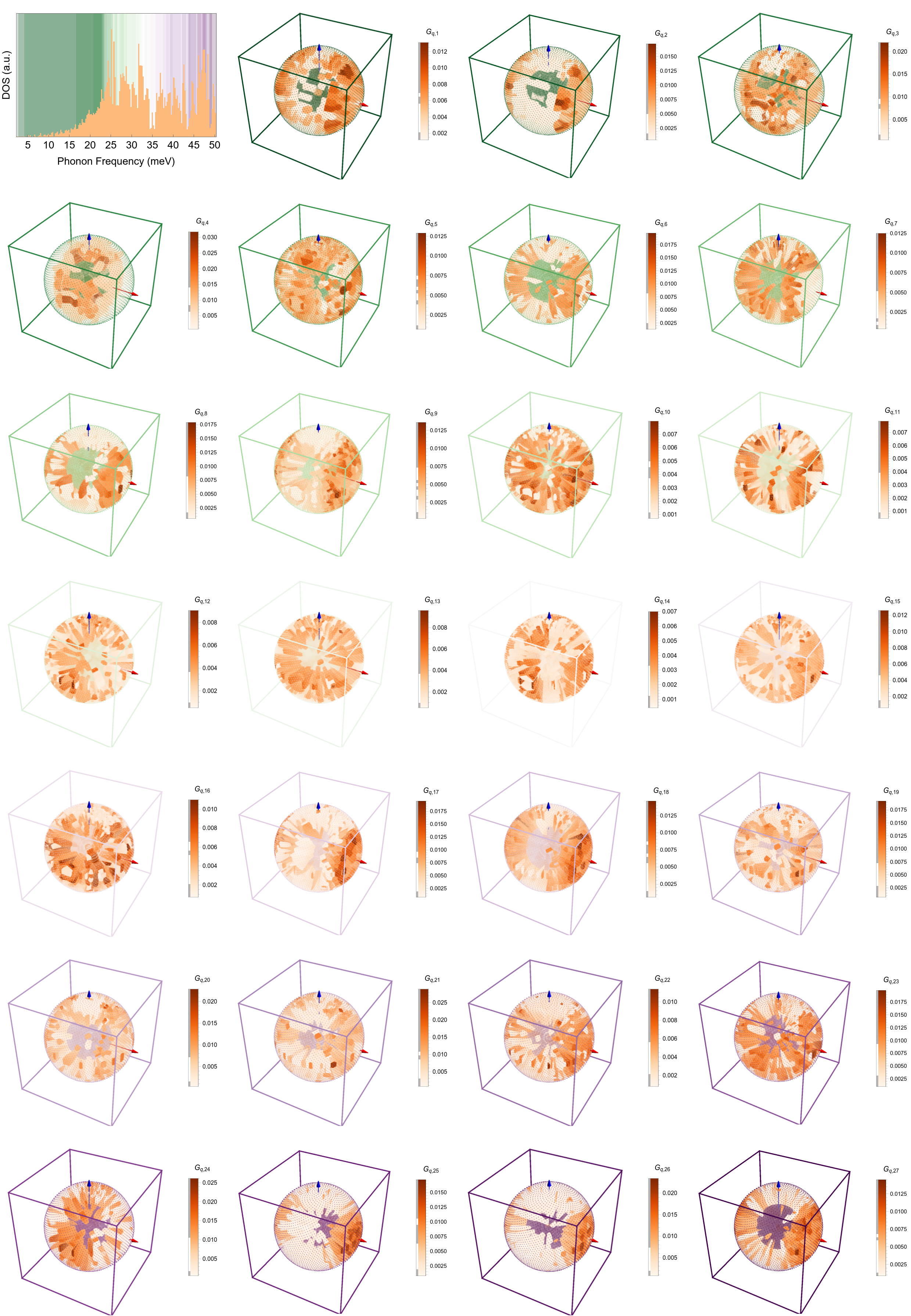}
    \caption{Phonon density of states (orange) of NbSi$_2$ (top left panel), together with mode-resolved Fermi-surface averaged electron-phonon coupling strength (the energy window for each mode is given by the color-coded background rectangles in the top left panel from low (green) to high (purple), also reflected in the box frame). 
    Subsequent panels show the angular-dependence in the phonon momentum $q$ of the Fermi-surface averaged electron-phonon coupling strength $G$ for various phonon modes $\alpha$, inside a spherical shell given by the mode's energy window. 
    The color-coding of the vectors represents the magnitude of $G$.
    In contrast to NbGe$_2$ (Fig.~\ref{fig:G_eph_NbGe2}), electron-phonon coupling appears to be largely isotropic in NbSi$_2$}
    \label{fig:G_eph_NbSi2}
\end{figure*}

We evaluate the phonon-mediated electron-electron scattering time as a Fermi-surface averaged version of Fermi's golden rule with a second-order electron-phonon matrix element according to 
\begin{equation}
\begin{split}
\frac{1}{\tau^{\text{ph}}_{\text{e-e}}}=\frac{2\pi}{\hbar \,g\left(\varepsilon_{F}\right)} \sum_{\alpha} \int \frac{\Omega\,d \mathbf{q}}{(2 \pi)^{3}} G_{\mathbf{q}\alpha}^2 \gamma\left(\beta \hbar \bar{\omega}_{\alpha\mathbf{q}}\right) \,,
\end{split} 
\label{eq:tau_eph}
\end{equation}

\noindent where $\beta^{-1}=k_BT$, the Boltzmann constant $k_B$ times the temperature $T$, and $\bar{\omega}_{\alpha\textbf{q}}=\omega_{\alpha\textbf{q}}(1+i\pi G_{\textbf{q}\alpha})$. The Fermi-surface averaged electron-phonon interaction strength can be obtained from
\begin{equation}
G_{\mathbf{q}\alpha}=\sum_{a b} \int \frac{g_{s} \Omega\,d \mathbf{k}}{(2 \pi)^{3}}\left|g_{a\mathbf{k},b\mathbf{k}+\mathbf{q}}^{\alpha\mathbf{q}}\right|^{2} \delta\left(\varepsilon_{a\mathbf{k}}-\varepsilon_{F}\right) \delta\left(\varepsilon_{b\mathbf{k}+\mathbf{q}}-\varepsilon_{F}\right)\,,
\label{eq:geph}
\end{equation}
and we introduce the function $\gamma(x)$: 
\begin{equation}
    \gamma(x) \equiv \int_{-\infty}^{\infty} \frac{dy}{\operatorname{sinch}^{2}\left(\frac{y}{2}\right)|x-y|^{2}}\,.
\end{equation}
We show $G_{\mathbf{q}\alpha}$ for both NbGe$_2$ (Fig.~\ref{fig:G_eph_NbGe2}) and NbSi$_2$ (Fig.~\ref{fig:G_eph_NbSi2}). Besides the orders of magnitude difference in the coupling between the two materials, it is remarkable that NbGe$_2$ shows a strong correlation between the direction of the phonon momenta and the coupling strength. In particular, modes that can contribute to electron scattering (in particular modes $\alpha=12,\ldots,15$) with a change in momentum along the in-plane axis (out-of-plane axis) show particularly strong (weak) coupling, providing a partial explanation of the pronounced anisotropy in the resistivity (cf.~Fig.~\ref{fig:scattering}(a)) compared to NbSi$_2$.

\subsection*{Hydrodynamic phase diagram}
We solve the electronic Boltzmann transport equation numerically, in the presence of momentum-conserving scattering. 
At steady-state, the evolution of the distribution function $f(\boldsymbol{r},\boldsymbol{k})$ for non-equilibrium electrons in the neighborhood of position $\boldsymbol{r}$ with wave vector $\boldsymbol{k}$ is given by:
\begin{align}
    \boldsymbol{v}_{\boldsymbol{k}} \cdot \nabla_{\boldsymbol{r}} f(\boldsymbol{r},\boldsymbol{k}) + e \boldsymbol{E} \cdot \nabla_{\boldsymbol{k}} f(\boldsymbol{r},\boldsymbol{k}) = \Gamma \left[f\right],
\label{eq:BTE}
\end{align}
where $\boldsymbol{v}_{\boldsymbol{k}}$ is the electron group velocity, and $\Gamma\left[f\right]$ is the collision integral.
We consider flow through a 2D channel ($\boldsymbol{r}=(x,y)$) of width $W$, for a spherical Fermi surface ($\boldsymbol{v}=v_{\mathrm{F}}(\cos{\theta}, \sin{\theta})$), at the relaxation time approximation level.
Under these approximations, we can linearize~\ref{eq:BTE}, to give an integro-differential equation in terms of an `effective' mean free path, $l_{\mathrm{eff}}$~\cite{molenkamp-hydro,wte2-hydro}:
\begin{align}
   \sin(\theta) \partial_y \, {l}_{\mathrm{eff}}(y,\theta) + \frac{{l}_{\mathrm{eff}}(y,\theta)}{l} = 1 + \frac{\tilde{l}_{\mathrm{eff}}}{l_{\mathrm{mc}}},
\label{eq:BTE-hydro}
\end{align}
where we use Mathhiessen's rule $l^{-1} = l_{mc}^{-1} + l_{mr}^{-1}$ to combine the depopulation of electronic states with a combined mean free path $l$.
Momentum-conservation is accounted for by the last term, which defines the `average' mean free path $\tilde{l}_{\mathrm{eff}}(y)$ and is directly proportional to current density, $j_x(y)$:
\begin{align}
    \tilde{l}_{\mathrm{eff}}(y) &= \int_0^{2\pi} \frac{d \theta}{\pi} \cos^2(\theta) l_{\mathrm{eff}}(y,\theta) \\
    j_x(y) &= \left(\frac{m}{\pi \hbar^2}\right) \varepsilon_{\mathrm{F}} e^2 \frac{E_x}{m v_{\mathrm{F}}} \tilde{l}_{\mathrm{eff}}(y).
    \label{eq:BTE-4}
\end{align}
where $m$ is the electronic effective mass, $E_x$ is the electric field component in the $x$ direction, and $v_F$ is the Fermi velocity. Equation~\ref{eq:BTE-hydro} is solved numerically by transforming it into a Fredholm integral equation of the 2\textsuperscript{nd} kind following Refs.~\cite{molenkamp-hydro,wte2-hydro}.

\subsection*{Computational details}
For the electron-phonon coupling, lifetime and resistivity calculations, we first obtain the electronic structure of each group V transition metal ditetrelide from first principles using density functional theory (DFT) as implemented in JDFTx~\cite{JDFTx}. 
We use fully relativistic optimized norm-conserving Vanderbilt pseudopotentials~\cite{VanSetten2018,Garrity2014,Hamann2013} for NbSi$_2$, NbGe$_2$, VSi$_2$, and VGe$_2$ and fully relativistic Rappe-Rabe-Kaxiras-Joannopoulos ultrasoft (RRKJUS) pseudopotentials~\cite{RRKJUS1990,DalCorso2014} for TaSi$_2$ and TaGe$_2$, all for the PBEsol exchange-correlation functional~\cite{Perdew2008}. 
For all compounds, we set a uniform $14\times 14\times 10$ $k$-point mesh across each 9-atom standard primitive unit cell, plane-wave energy cutoffs of 41 Hartrees, and Marzari-Vanderbilt (``cold'') smearing~\cite{Marzari1999} with a $0.001$ Hartree width. 
The lattice constants and ionic positions of each material are relaxed such that the forces on all atoms are less than $10^{-9}$ Ha/Bohr. 
We employ a frozen phonon approximation with a $2\times 2\times 2$ supercell to calculate each set of phonon states, and construct bases of 78 maximally localized Wannier functions (MLWFs)~\cite{MarzariVanderbilt1997,Souza2001} from the DFT plane-wave electronic states, which allow us to accurately interpolate electron and phonon properties to denser corresponding $\mathbf{k}$ and $\mathbf{q}$ meshes for effective Brillouin zone integration~\cite{Giustino2007}. 
For the Coulomb-mediated electron-electron scattering lifetimes of NbGe$_2$ and NbSi$_2$, dielectric matrix cutoffs of 200~eV were used with an energy resolution of 0.1 eV.

\subsection*{Kohn-Sham bands and phonon spectra in transition metal ditetrelides}
\begin{figure*}[t]
    \centering
    \includegraphics[width=\textwidth]{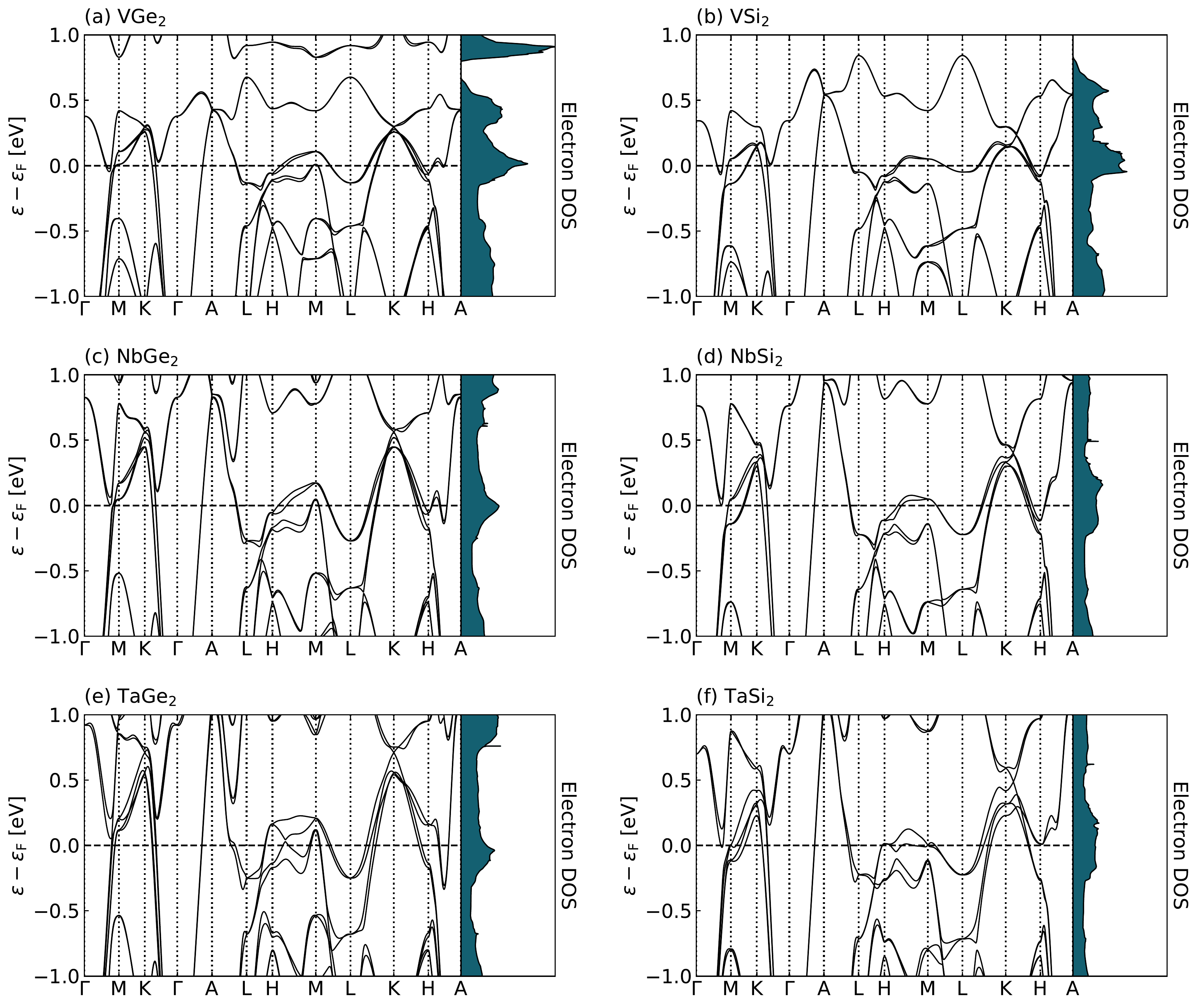}
    \caption{\textit{Ab initio} electron band structure and density of states close to the Fermi level for all six transition metal ditetrelides that crystallize in space group 180 or 181.}
    \label{fig:electrons}
\end{figure*}

\begin{figure*}[t]
    \centering
    \includegraphics[width=\textwidth]{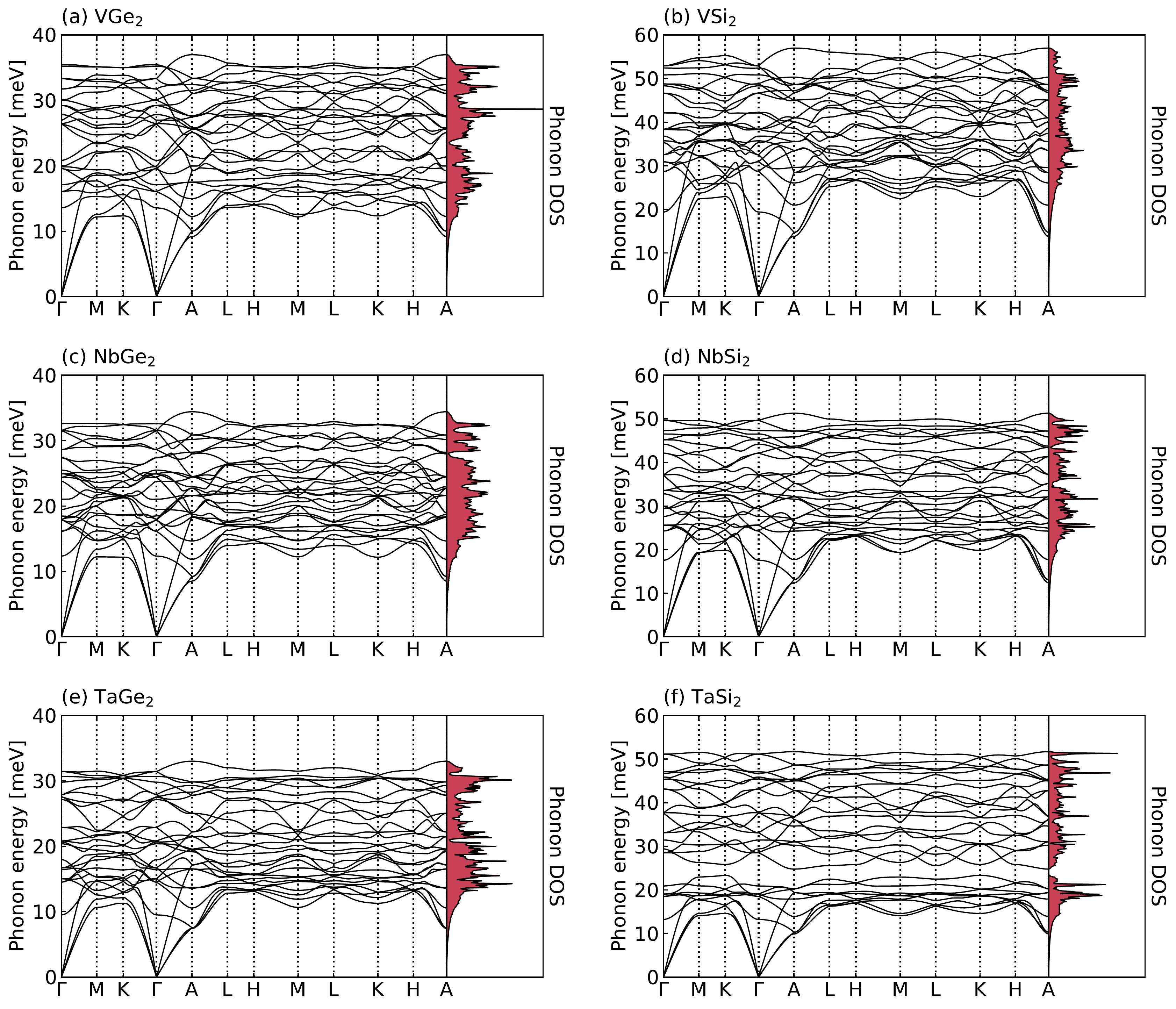}
    \caption{\textit{Ab initio} phonon spectra and density of states for all six transition metal ditetrelides that crystallize in space group 180 or 181.}
    \label{fig:phonons}
\end{figure*}

To complement the main text, we show the single-electron band structure (Fig.~\ref{fig:electrons}) obtained as outlined above, as well as the phonon spectra and their density of states (Fig.~\ref{fig:phonons}) for all six transition metal ditetrilides that crystallize in space group 180 or 181.

\end{document}